\documentclass[manuscript,nonacm]{acmart}
\renewcommand\footnotetextcopyrightpermission[1]{}
\AtBeginDocument{%
  }

\def\repocount{100,356 }

\copyrightyear{2026}
\acmYear{2026}
\setcopyright{cc}
\setcctype{by}
\acmConference[MSR '26]{23rd International Conference on Mining Software Repositories}{April 13--14, 2026}{Rio de Janeiro, Brazil}
\acmBooktitle{23rd International Conference on Mining Software Repositories (MSR '26), April 13--14, 2026, Rio de Janeiro, Brazil}
\acmPrice{}
\acmDOI{10.1145/3793302.3793305}
\acmISBN{979-8-4007-2474-9/2026/04}

\begin{document}

\title{HackRep: A Large-Scale Dataset of GitHub Hackathon Projects}


\author{Sjoerd Halmans}
\email{s.halmans@student.tue.nl}
\affiliation{%
  \institution{Eindhoven University of Technology}
  \city{Eindhoven}
  \country{The Netherlands}
}

\author{Lavinia Paganini}
\email{l.f.paganini@tue.nl}
\affiliation{%
  \institution{Eindhoven University of Technology}
  \city{Eindhoven}
  \country{The Netherlands}
}
\author{Alexander Serebrenik}
\email{a.serebrenik@tue.nl}
\affiliation{%
  \institution{Eindhoven University of Technology}
  \city{Eindhoven}
  \country{The Netherlands}
}
\author{Alexander Nolte}
\email{a.u.nolte@tue.nl}
\affiliation{%
  \institution{Eindhoven University of Technology}
  \city{Eindhoven}
  \country{The Netherlands}
}
\affiliation{%
  \institution{Carnegie Mellon University}
  \city{Pittsburgh}
  \state{PA}
  \country{USA}
}

\renewcommand{\shortauthors}{Halmans et al.}

\begin{abstract}


  Hackathons are time-bound collaborative events that often target software creation.
  Although hackathons have been studied in the past, existing work focused on in-depth case studies limiting our understanding of hackathons as a software engineering activity.

  To complement the existing body of knowledge, we 
  introduce HackRep, a dataset of \repocount hackathon GitHub repositories. 
  We illustrate the ways HackRep can benefit software engineering researchers by presenting a preliminary investigation of hackathon project continuation, hackathon team composition, and an estimation of hackathon geography.
  We further display the opportunities of using this dataset, for instance showing the possibility of estimating hackathon durations based on commit timestamps.

\end{abstract}





\maketitle

\section{Introduction}
Hackathons are time-bounded events during which individuals form teams to collaborate on a project that is of interest to them~\cite{10666667}. The most common goal of hackathons is developing software. This can be for instance creating tools or expanding them.
Hackathons are not limited to the computer science domain, though, with other examples being citizen science and higher education~\cite{amefon2022,gama2018hackathons,taylor2018everybody}.
Hackathon goals may also vary, like creating software, knowledge sharing or even entrepreneurship~\cite{10666667}.

Currently, few large-scale public datasets exist to perform research on hackathons, with many studies focusing on in-depth case studies. 
Existing quantitative work in turn is based on smaller datasets that include specific types of events, like the Devpost dataset used by several papers which mainly includes collegiate events~\cite{Nolte,Mahmoud,falk2025hackathons}. 
While in-depth case studies can be useful, their focus on individual events in specific domains can attract a specific audience. 
This in turn harms our ability to transfer findings beyond the specific study contexts.
This, in addition to hackathons growing popularity, illustrates the need of a large-scale public dataset to enable quantitative future research into a wide variety of hackathons.

To achieve this goal, we present HackRep, a dataset containing \repocount  hackathon projects mined from GitHub.
By leveraging the domain-agnostic nature of GitHub, in addition to the size of the dataset, we allow future research to be conducted on a quantitative scale, with the potential for result to be transferable ~\cite{lincoln_guba_1985_trustworthiness} across domains. We illustrate the usefulness of the dataset through several research questions that may be answered with the use of HackRep, with the goal of inspiring future work.
The HackRep dataset and the scripts used to create it are publicly available at \href{https://doi.org/10.5281/zenodo.17572684}{Zenodo}\footnote{https://doi.org/10.5281/zenodo.17572684}.

\section{Related works}
While there is a lack of a wide range of public datasets, the study of hackathons is not new~\cite{Falks}, with studies focusing on several different aspects of hackathons.
Existing work focuses on understanding hackathons themselves, studying participant collaboration, inclusivity, hackathons as a platform, and much more.
For instance, Hope et al. ~\cite{Hope-Participatory} explore hackathons as design spaces and study how to improve inclusivity.
Meanwhile, Cobham et al. ~\cite{Cobham-Enterprise} analyse hackathons as a platform for entrepreneurship among students.

In parallel, other studies focus on the impact of hackathons in terms of education, continuation, and other aspects.
Studies by Nolte et al. ~\cite{Nolte, Nolte_corp} investigate the continuation of hackathon projects, finding that short-term and long-term continuation are different phenomena.
In the same domain, works by Imam et al. ~\cite{Imam-Code, Imam-Tracking} study the creation and reuse of hackathon code, furthering our understanding of the evolution of software code beyond the hackathon it was conceived in.
Other studies in this context explore hackathon contributions to education and collaboration ~\cite{Busby-Gaps, Falks, falk2025hackathons}.

While existing studies -- like the ones mentioned before -- provide important insights into hackathons and their impact, their findings are difficult to transfer beyond the contexts they were obtained in.
Some studies aim to go beyond specific events ~\cite{Nolte, Imam-Code, Imam-Tracking}, but still have limitations due to the use of a largely domain specific dataset.
Thus, a large-scale dataset containing hackathons from a variety of domains, would be beneficial for continued research.

\section{Methodology}
To enable research into hackathons, the dataset should include information about events, teams, projects, and individuals, using~\cite{Nolte} as a basis.
Thus, to construct the dataset, we followed a 3-step process: \textbf{Data collection} (section \ref{subsec:data}), \textbf{Filtering} (section \ref{subsec:filtering}), and \textbf{Enrichment} (section \ref{subsec:Enrichment}). 
Figure \ref{fig:overview} shows an overview.

\begin{figure}[h!]
    \centering
    \includegraphics[width=0.45\linewidth]{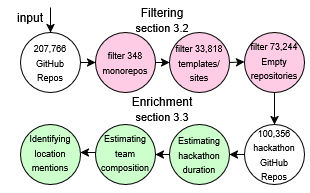}
    \caption{Data filtering and enrichment workflow}
    \label{fig:placeholder}
\end{figure}

\subsection{Data collection}
\label{subsec:data}
In order to facilitate quantitative research on a large scale while staying domain-agnostic, it is important to include a wide variety of hackathon projects. 
As we are focusing on software projects, GitHub is the most suitable source to mine projects from as it includes a wide variety of repositories.
We first collected the URLs of all repositories containing the keyword "hackathon". This appears suitable since it is the most widely used term for such events ~\cite{Imam-Code} and existing work confirms that other keywords only yield minimal gains~\cite{10666667}.

This query yielded 283,264 repositories in total.
Our goal is to create a dataset that can be used for research on modern day hackathons. Work by Falk et al.~\cite{falk2025hackathons} has indicated that hackathons change rapidly over the years and adapt to societal and technological developments.
Thus, the decision was made to limit our dataset to data spanning over the last five years. Omitting any projects older than this, we are left with 206,421 GitHub repositories.
In an attempt to find additional projects, we also queried for projects containing the topic "hackathon".
This additional search yielded 4868 repositories, which, after filtering duplicate projects, brings the total project count to 207,766 GitHub repositories.
\subsection{Filtering}
\label{subsec:filtering}
Our goal is to create a dataset of hackathon projects. 
However, upon inspection of the collected data, we realized other types of repositories were also included.
We identified the following types of repositories:

\begin{itemize}
  \item \textbf{Collection}: The repository is used as an overarching repo for multiple or perhaps all repositories created during a hackathon. 
  \item \textbf{Individual}: The repository is used to store a single project that participants worked on during a hackathon, either maintained by a team of participants or an individual participant.
  \item \textbf{Template/website}: The repository contains template code that can be used to start a hackathon. This can be useful, when a certain hackathon has a long setup time, but you wish to focus on the deeper implementation and not waste time on setting up projects. A similar version of these are repositories that consist of only a website.
\end{itemize}

Since our goal is to create a dataset of hackathon projects, templates, websites, and collections should not be included.
Our first step of identifying and removing such projects is the removal of all collections from the dataset using a keyword search.
Specifically, all repositories containing the words 'monorepo' or 'umbrella repo' were removed from the dataset, removing 348 repositories. 
These were chosen after a small inspection indicated collection repositories often contained either word in their README files.
While few collections may remain, we elected to not filter further to avoid losing hackathon repositories, as many collection repositories share characteristics with individual repositories.
Thus, increased filtering in this step would have likely resulted in the removal of a large amount of individual repositories.
Secondly, templates and websites are either not hackathons, or not in the form that we wish to collect. We thus decided to also remove these projects from the dataset.
We filter these projects with the use of the data on project languages as indicated by GitHub, in this case, CSS and HTML.
By omitting all projects from the dataset with these languages as their main language, 33,818 projects are removed. 

After these filtering steps, 173,600 repositories remained in the set, leaving us with our final step in the filtering process.
A GitHub repository can contain information in various ways, however, we argue that without a README or commits, a repository provides limited possibilities for further investigation. 
To back this claim, we performed a randomized inspection of a portion of the remaining repositories to identify if README files and commits contain significant information. 
During this inspection, it became clear that most projects lacking a README file or commits did not offer meaningful information in other locations (for instance issues).
We thus remove all projects from our dataset that either contained 0 commits, or did not contain a README file, omitting 73,244 repositories.
After these filtering steps, \repocount repositories remain.
\subsection{Enrichment}
\label{subsec:Enrichment}
At this stage, we have a set of repositories related to hackathons.
To allow increase the usefulness of the dataset for research, however, more information is required. 
To this end, we add information on hackathon duration, project contributors, and estimated locations, and finally attempt to identify scientific hackathons in the dataset. The process is explained in the following subsections.

\subsubsection{\textbf{Hackathon duration}}
To study hackathon continuation, team composition, or even individual hackathon participation, knowing the start and end of a hackathon is crucial.
To this end, we estimated hackathon start and end dates using burst detection on commit timestamps.
To implement this, we made use of Kleinbergs burst detection algorithm ~\cite{Kleinberg2003} to detect active time frames of hackathons. The algorithm takes as input, a list of timestamps and hyperparameters. 
It then identifies time frames in which relatively many timestamps occur, which we then marked as a 'burst' and assigned a level based on intensity.
These bursts can then be interpreted in a variety of ways.
One assumption made here, is that the first burst of activity indicates the hackathon runtime. 
We confirmed this by manually inspecting 30 repositories, and confirming that first bursts were indeed during the hackathon duration, being in line with findings from previous work \cite{Nolte_corp}

This was done with the knowledge that some projects may have been active before their hackathon had commenced. 
However, due to the nature of hackathon projects, the assumption that these are minimal outliers was made.

\subsubsection{\textbf{Team composition}}

In addition to duration, identifying contributors during the hackathon duration can be valuable for studies focusing on individual performance or team composition.
To identify individuals that contributed to a hackathon, we create a list per repository of all contributors that created a commit during the hackathon timeframe.

\subsubsection{\textbf{Location mentions}}
Often, studies may mention the location in which their research took place, this can be useful as it may provide information regarding potential participants.
Currently, most datasets used in hackathon research are primarily focused in Europe and North America.
To identify the hackathon location, we utilized the text provided in the README as this is the mostly likely place for this information to be present/
To this end, we made use of the Geotext ~\cite{geotext} python library to identify locations inside of README texts.

\subsubsection{\textbf{Manual annotation}}
\label{sec:manual}
The collection of large quantities of text allows for research using language processing to identify certain types of repositories.
Due to their low representation in datasets like Devpost, we were particularly interested in identifying and investigating scientific hackathons.
We chose to create a classification model to identify them on a large scale.
In addition, this process serves as a small-scale validation of the gathered data, ensuring that a significant number of gathered repositories is non-trivial and is suitable for future research.
To this end, the main author manually annotated a subset of 573 repositories as scientific or not to serve as training data for the model, using the following criteria:

\begin{itemize}
    \item The goal of the hackathon linked to the repository or repository itself is to contribute to scientific research.
    \item The repository is in itself a paper.
    \item The repository is used in scientific research (published papers) and links to specific papers.
\end{itemize}

By following these criteria, we attempt to reduce the effects of bias present due to a single author labeling these repositories.
These specific criteria were chosen as they are easily identifiable by manual inspections (for instance finding paper references in README files) and they leave little to no room for ulterior interpretation.
Subsequently, they were manually classified with one of three labels during the annotation process (Yes, No, and Maybe).


A small script was used to manually annotate these repositories, giving the author one repository to inspect at a time by using the GitHub link.
Apart from README inspections, we also identified the hackathons events that projects were linked to where possible, using direct mentions in the README or description.
From these observations, the main author marked each investigated repository with either Yes, No, or Maybe in a separate column of the dataset. These classifications were later used to train our machine model. In addition to inspecting README files for potential identifiers for scientific hackathons (for instance, links to papers, or the professions of collaborators), linked hackathons were also identified where possible. 
This possibility exists because some repositories include the URL link to the hackathon they were associated with in the description or README, allowing for manual inspection.
Furthermore, some repositories included the names and year of the hackathons, allowing the author to find these hackathons online during the annotating process.
\begin{figure*}[htbp]
    \centering
    \begin{minipage}{0.45\textwidth}
        \centering
        \includegraphics[width=\linewidth]{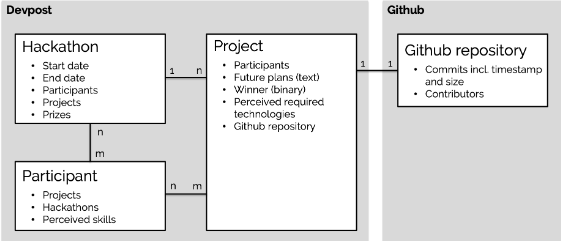}
        \caption{Variables contained in the Devpost dataset~\cite{Nolte}}
        \vspace{-6pt}
        \label{fig:devpost}
    \end{minipage}\hfill
    \begin{minipage}{0.45\textwidth}
        \centering
        \includegraphics[width=\linewidth]{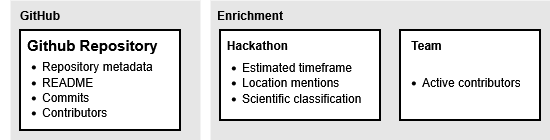}
        \caption{Variables contained in the HackRep dataset}
        \vspace{-6pt}
        \label{fig:overview}
    \end{minipage}
\end{figure*}

\subsubsection{\textbf{Automated classification}}
Making use of the manually annotated subset of GitHub repositories, we trained a logistic regression model to identify scientific repositories using the following fields.
This binary classification model uses of a composite text field, including the following information:
\begin{itemize}
    \item[] \textbf{name} - The name of the repository
    \item[] \textbf{full\_name} - Full name of the repository link
    \item[] \textbf{description} - A short description of the repository
    \item[] \textbf{topics} - A list of topics related to the repository
    \item[] \textbf{README} - The README file located in the repository
\end{itemize}
This field is then turned into a feature space using a Term Frequency-Inverse Document Frequency Vectorizer (TF-IDF). 
This allows the model to recognize word patterns that may be linked to scientific projects. 
The first round of automated classifying had an accuracy score of 77\%. 
Notably, the minority class (scientific repositories) had an F1 score of 63\% in comparison to the 84\% of the majority class. 
This first round classified 694 repos as scientific, and 99662 as non-scientific. 
In addition, scientific repositories are less common than non-scientific repositories according to manual inspection of the dataset. 
We then ran a second manual annotation round, focusing on repos with a high likelihood of being science-based on initial ML results. 
After concluding this additional round of annotating, we came to a total of 573 repositories manually annotated.
With these additional annotations we ran the classifier for another round.
In total, we classified 449 as scientific, and 99,907 as non-scientific with an F1 score of 69\%.

\section{Dataset structure and findings}
For this dataset, we have gathered \repocount GitHub repositories of hackathon projects with their README files and commits.
In addition, we have enriched the dataset with the following estimations: duration, contributors, scientific-focus, and geographical mentions.

\begin{table}[h!]
\vspace{-.1in}
  \centering
  \caption{Enrichment statistics}
  \label{tab:enrichment}
  \begin{tabular}{r r r r}
    \toprule
    Repositories & avg. duration & avg. team size & \% with locations \\
    \midrule
    \repocount & 36 hours & 3 participants & 17.7\%  \\
    \bottomrule
  \end{tabular}
\end{table}

Table \ref{tab:enrichment} provides an overview of the enriched dataset. 
We observe that the average estimated duration in our set is approximately 36 hours, consistent with prior findings that hackathons typically span a couple of days.
Furthermore, we note that only 17.7\% of projects have mentions of locations in their README.
In terms of languages, we find that scripting languages seem to be most prevalent in our dataset's projects (Table~\ref{tab:annotation‐rounds}). 
This aligns with the goal of swift software creation, which these languages excel in.

\begin{table}[h!]
\vspace{-.1in}
  \centering
  \caption{Prominent languages}
  \label{tab:annotation‐rounds}
  \begin{tabular}{l r r r r}
    \toprule
    Languages & JavaScript & Python & TypeScript & Jupyter \\
    \midrule
    Repo count & 27,825 & 20,714 & 18,501 & 14,449 \\
    \midrule
    Percentage & 27.73\% & 20.64\% & 18.44\% & 14.40\% \\
    \bottomrule
  \end{tabular}
\end{table}

In terms of identifying scientific hackathons, we note that our machine model classifies a few repositories as scientific. 
During manual inspection, we found that those marked fit our perception of scientific hackathons, which may illustrate that a more in-depth NLP analysis may be able to identify a larger number of scientific hackathons in our dataset. We finally note that team size averages at three participants, this value was seen to have various outliers, potentially due to continued projects, or large-scale hackathons on one side, and solo hackathon projects on the other side.

\section{Discussion and Future Work}
\label{sec:future}
By creating HackRep, we enable further research into hackathons on a larger scale.
To illustrate this, we offer a few examples of research questions that could be answered using HackRep:

\begin{itemize}
    \item Can studies performed using a case study or smaller domain-focused dataset ~\cite{Nolte, Imam-Code, Imam-Tracking} be replicated to transfer their results to broader contexts?
    \item What are the results of scientific hackathons, and does their work get continued beyond their hackathon?
    \item What is the quality of software code created during hackathons?
    \item Do forked hackathon projects often lead to continuation?
\end{itemize}

These are but a few examples of questions that may be answered using the HackRep dataset, and further enrichment 
can expand these possibilities even further. In addition, we note that the projects in our dataset (Figure \ref{fig:overview}) differ from those in the non-public Devpost dataset (Figure \ref{fig:devpost}). After removing projects without a GitHub repository from the Devpost dataset, we were left with 82925 projects to compare with our dataset. The datasets share 2307 projects, indicating minimal overlap.
In addition, HackRep includes projects across a wide range of domains, setting it apart from the Devpost dataset, which primarily contains collegiate projects.
\section{Threats to validity}
\textbf{Single annotator bias}
Bias may be present in the annotations due to a single author creating them. We argue, however, that by following well-defined criteria, this bias is minimized.
\\
\textbf{Keyword filtering}
Several parts of the filtering in this project used keywords, which may have led us to remove scientific hackathon projects.
However, since our goal is to allow further research, we focused on classification accuracy over including potentially irrelevant repositories.
\\
\textbf{Automated classification}
The use of automated classification could introduce bias. 
To counteract this, an attempt was made to create a suitable ground truth to train the model on. 
\\
\textbf{Buildability \& dependency compatibility} 
During this study, no repositories were inspected to confirm whether the code they contained worked. 
In addition, we abstained from confirming that the gathered code compiles due to the constant evolution of code.

\section{Conclusion}
We present HackRep: \repocount hackathon code repositories enriched with metadata, commit-level features, and README text mining.
The dataset enables researchers to conduct quantitative, large-scale analyses of hackathon-related activity, supporting investigations into topics such as project continuation, scientific engagement, or temporal collaboration patterns. 
Future extensions may include linking additional data sources, integrating social or network-based features, and expanding temporal coverage to capture longitudinal dynamics in hackathon participation and output.

\begin{acks}
This publication is part of the project ``A new era of room acoustics simulation software: from academic advances to a sustainable open-source project and community'' with file number 19430 of the research programme Open Technologieprogramma (OTP), which is (partly) financed by the Dutch Research Council (NWO). 
\end{acks}

\newpage
\bibliographystyle{ACM-Reference-Format}
\bibliography{biblio}

\appendix

\end{document}